\documentclass[debug,overfull]{epl}
\usepackage{epsf}
\def\ltsim{\vbox {\hbox{\lower .8\baselineskip \hbox{$<$}} \break
                 \hbox{\lower 0.2\baselineskip \hbox{$\sim$}} } }
\def\k{{\bf k}}

\def\d{{\bf d}}
\title{Non--Linear Transport through a Molecular Nanojunction}
\shorttitle{Molecular Nanojunction}
\author{
Matthias H. Hettler\inst{1}, Herbert Schoeller\inst{2} and 
Wolfgang Wenzel\inst{1}}
\institute{
\inst{1} Forschungszentrum Karlsruhe, Institut f\"ur Nanotechnologie, 
Postfach 3640, D-76021 Karlsruhe, Germany\\
\inst{2} RWTH Aachen, Theoretische Physik A, D-52056 Aachen, Germany}
\begin{document}
\maketitle
\begin{abstract}
  We present a simple model of electrical transport through a 
  metal--molecule--metal nanojunction
  that includes charging effects as  well as   aspects of the electronic 
  structure of the molecule. The interplay  of  a large charging energy and 
  an asymmetry of the metal--molecule
  coupling can lead to various effects in  non--linear electrical  transport.
  In particular, strong negative differential conductance is observed
  under certain conditions. 
\end{abstract}
{\it Introduction.}  Single molecule electron transistors
offer exciting perspectives for further minituarisation of
electronic devices with a potentially large impact in applications.
To date several experiments have shown the possibility to attach
individual molecules to leads and to measure the electrical transport.
Two terminal transport through a single molecule
\cite{reed-etal,kergueris-etal,porath2-etal,reichert-etal} or other 
nanoscale objects \cite{bezryadin-etal,klein-etal,persson-etal} has been 
achieved by deposition of the object between two fixed electrodes or a
conducting-tip STM above an object attached to a conducting substrate
\cite{bumm-etal,banin-etal,alperson-etal,porath1-etal}. Interesting
and novel effects, such as negative differential conductance (NDC), were
observed in one experiment \cite{chen-etal}, which still 
needs satisfactory theoretical explanation.

Several factors are important for single-molecule
transport: For nanoscale objects the capacitance $C$ is very small.
Consequently, the energy to charge (or uncharge) the molecule 
$E_C=e^2/2C$ can be very large, of the order of electron volts. This leads
to the phenomenon of Coulomb blockade and makes room temperature
single-electron transistors (SET) based on such molecules a distinct
possibility. In contrast to SETs based on
metallic islands \cite{grabert-devoret,sohn-etal}, 
molecular devices have a more complicated electronic structure that, 
in principle, can be chemically
'designed'. Gaps in the I--V curve are not only determined by $E_C$ but
predominantly by the structure of the molecular bands
\cite{kergueris-etal,porath2-etal}. Therefore, it is important to
consider the interplay of charging effects with the specific structure
of the molecular orbitals. For electronic transport we will see
that, in particular, the specific {\it spatial} structure 
of the molecular orbitals is crucial.

In this letter we study the impact of such a low-energy
electronic structure on electronic transport and demonstrate that it
can result in non--trivial conductance under reasonable generic
assumptions.  A full quantitative treatment of a molecule
in contact with metal electrodes including many--body interactions on
the molecule and strong molecule-electrode coupling is still out of
reach.  In this paper, we therefore study a simple model for a
'generic' molecule that can be derived in principle from full
electronic structure calculations and that includes capacitive
interactions within the Coulomb blockade model as well as absorption
and emission of photons.  In contrast to most other theoretical
work \cite{mujica-etal,datta-etal,magoga,emberly-etal,ventra-etal}, we
assume weak coupling between electrodes and the functional part of the
molecule, a scenario that can be realized by proper design of molecule
and contacts. We predict that the
interplay of charging effects, the spatially non--trivial electronic
structure and resulting asymmetric coupling to the electrodes leads to
interesting phenomena in non--linear transport.We find that the I--V
curve can show strong NDC behaviour and thus
may provide a possible mechanism to explain recent
experiments \cite{chen-etal} showing similar NDC behaviour.  Finally,
we give a specific example that displays the required 
electronic structure in a simple aromatic molecule.

{\it The Model.}  In the low-voltage regime only a few of the
molecular levels will contribute to transport. For the purpose of this
paper, we assume without loss of generality that there are only two
participating molecular levels that are both unoccupied at zero
voltage, which we designate as LUMO and LUMO$+1$. In particular in
$\pi$ electron systems of aromatic molecules 
it is relatively easy to realize a
situation, where two closely spaced MOs are separated far
from both from the HOMO and the other LUMOs.  The other
MOs are assumed inert (always occupied or always empty), as we ignore
co-tunnelling effects.  

The MO Hamiltonian of the 'reduced' system can be written as:
\begin{eqnarray}
  {\cal H} & = & \sum_{i\sigma} \epsilon_{i\sigma}
c^\dagger_{i\sigma}c_{i\sigma} +
\sum_{ijkl\sigma\sigma^\prime} V_{ijkl} c^\dagger_{i\sigma}
c^\dagger_{j\sigma^\prime}c_{k\sigma^\prime} c_{l\sigma},
\label{eq:fullh}
\end{eqnarray}
where the operators $c_{i\sigma},c^{\dagger}_{i\sigma}$
destroy/create electrons with spin $\sigma$ in MO $i$.  
This Hamiltonian contains a large number of parameters even for the two
level system. In order to make contact with the quantum dot physics
and to elucidate similarities and crucial differences that occur in
single molecule transport it is useful to introduce an effective model
for the molecular Hamiltonian:
\begin{eqnarray}
\label{eq:redh} 
&& H_{mol} = \epsilon_1 N_1 + \epsilon_2 N_2 +  E_C\, (N_1 +N_2)^2  \\ 
&& + \frac{U}{2} \sum_l N_l (N_l - 1) 
+  \frac{\Delta_{ex}}{2}\,\sum_{\sigma,\sigma'} 
c^{\dagger}_{1\sigma}c^{\dagger}_{2\sigma'}c_{1\sigma'}c_{2\sigma}
,\nonumber
\end{eqnarray}
where $N_l$ is the occupation of the MO $l$. As unit of energy we take
$\Delta\epsilon=\epsilon_2-\epsilon_1$, the ``bare'' MO splitting. The
diagonal electronic repulsion terms of Eq.~(\ref{eq:fullh})
for the two MOs together with
capacitive interactions with the leads can be rewritten
as $E_C\, (N_1 +N_2)^2 + \frac{U}{2} \sum_l N_l (N_l - 1)$.  
(for simplicity, we assumed an orbital independent Hubbard-like 
repulsion U for double occupancy of a MO). 
$\Delta_{ex}$ is a Hund's rule triplet-singlet
splitting for the two electronic levels. In comparison to
Eq.~(\ref{eq:fullh}) we have neglected single electron hopping mediated
by two-electron screening effects (which are small for most $\pi$
systems in the chosen basis).

\begin{figure}
\centerline{\includegraphics[width=11cm]{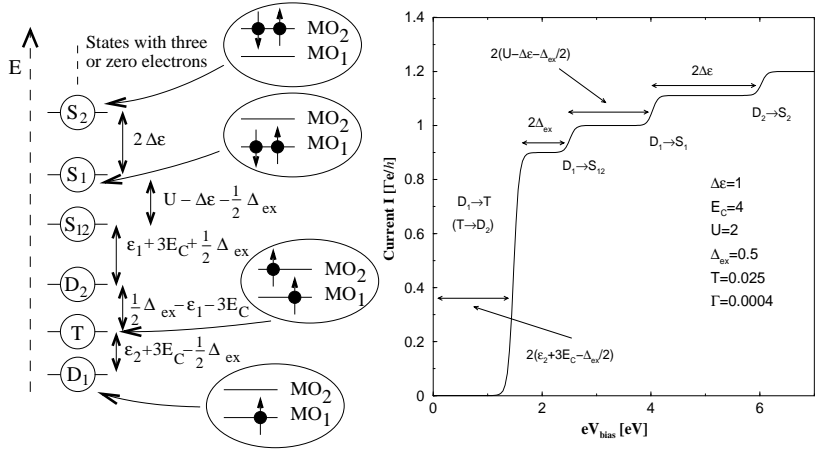}}
\caption {Equilibrium molecule energies for model 
Hamiltonian~(\protect\ref{eq:redh}) and resulting
current--voltage characteristics. We choose energies such that the
triplet states ($T$) lies below the singlets ($S_{12}$ $S_1$
$S_2$) and between the two possible doublets.
$\epsilon_1=-12$, $\epsilon_2=-11$, $(\Delta\epsilon=1)$, 
$U=2$, $\Delta_{ex}=0.5$, $E_C=4$, $\Gamma=0.004$, $T=0.025$.}
\label{en_ivnew}
\end{figure}

The bias is dropped symmetrically over the molecule, so no capacitive
shifts of energies appear (this can be easily included).  A likely 
energetic `term scheme' for the single- and two-particle states
described by model Hamiltonian Eq.~(\ref{eq:redh}) is depicted
in the left panel of Fig.~\ref{en_ivnew}.
The electrodes are considered as non--interacting electron reservoirs.
The reservoirs are assumed to be occupied according to an equilibrium
Fermi distribution function $f_\alpha(\omega)=f(\omega-\mu_\alpha)$,
where $\mu_\alpha$ denotes the electrochemical potential of electrode
$\alpha=L,R$.  The MOs couple to the electrodes via tunnelling contacts
with (possibly) very different coupling strength $t^{\alpha}_l$.
\begin{equation}
H_{mol-leads} =   ({\Gamma\over 2\pi\rho_e})^{1/2}
\sum_{\k\sigma\alpha l}\left( t^{\alpha}_l
  c^{\dagger}_{l\sigma}a_{\k\sigma\alpha} + h.c. \right)\;,
\end{equation}
where $\rho_e$ is the density of states (assumed constant)
of the non--interacting electrons in the leads, described by operators
$a_{\k\sigma\alpha}$. $\Gamma$ denotes the scale of the broadening of the 
MOs due to the coupling to the leads.
The dimensionless tunnelling parameters $t^{\alpha}_l$ can depend 
on $\alpha,l$. 
 
In order to investigate whether the predicted effects are robust, we
include a coupling of the molecule to a (broad band) boson field which
simulates the relaxation of excited states in a real molecule by
coupling of the electrons to an electromagnetic field (photons) and
vibrations (phonons)\cite{excitation}.

{\it Theoretical Approach.}         
We use a Master equation approach for the occupation 
probabilities $P_s$ of the molecular many--body states. The transition 
rate $\Sigma_{ss'}$ from state $s'$ to $s$ is computed  up to
linear order in $\Gamma$ using golden rule (second order perturbation
theory) in both the electrode-molecule and the bosonic coupling. 
For the transition rates we have
$\Sigma_{ss'}=(\sum_{\alpha,p=\pm}\Sigma_{ss'}^{\alpha p}) +\Sigma^b_{ss'}$
where $\Sigma_{ss'}^{\alpha p}$ is the tunnelling rate 
to/from electrode $\alpha$ for creation (p=+) or destruction
(p=-) of an electron  on the molecule. 
We have
\begin{equation}
\Sigma_{ss'}^{\alpha +}=\Gamma f_\alpha(E_s-E_{s'})
\sum_\sigma |\,\sum_l t^\alpha_l <s|c^\dagger_{l\sigma}|s'>\,|^2\;,
\end{equation}
and a corresponding equation for $\Sigma_{s's}^{\alpha -}$ by
replacing $f_\alpha\rightarrow 1-f_\alpha$.
The boson-mediated rates 
$\Sigma^b_{ss'}$ describe absorption and emission of bosons. For photons
we have
\begin{equation}
\Sigma^b_{ss'}=g_{ph} \frac{4 e^2}{3 \hbar^3 c^3}
(E_s-E_{s'})^3 N_{b}(E_s-E_{s'})
\, | <s|\d|s'>|^2\;,
\end{equation}
where $\d$ is the dipole operator and 
$N_{b}(E)$ denotes the equilibrium Bose function.
$g_{ph}$ is a parameter that allows us to modify the strength 
of the coupling to simulate increased dipole moment or other sources of
relaxation. $g_{ph}=1$, unless noted. This value
corresponds to a dipole of charge $e$ and length 1\AA.

We determine the $P_s$ by solution of the stationarity condition:
\begin{equation}
\dot{P_s}= 0 = \sum_{s'} (\Sigma_{ss'} P_{s'} - \Sigma_{s's} P_s)\;.
\end{equation}

The current in the left and right electrode can then be
calculated via 
\begin{equation}
 I_{\alpha}= e \sum_{ss'} (\Sigma^{\alpha +}_{ss'} P_{s'} -
\Sigma^{\alpha -}_{s's} P_s)\;.
\end{equation}
The bosonic transition rates do not contribute directly to the current,
since they do not change the particle number on the molecule.

{\it Results.}  
The effective Hamiltonian Eq.~(\ref{eq:redh}) affords several
generic scenarios for NDC. The NDC is generic in the sense that NDC
will occur at some bias for an initially charged molecule (case (1)) 
as well as an initially empty molecule (case (2)).

Case (1): The right panel of 
Fig.~\ref{en_ivnew} shows the I--V--characteristics for equal tunnelling
couplings $t^\alpha_l=1$ with a symmetric bias, $\mu_L=-\mu_R$. 
There are four characteristic steps which are related to the onset of 
the triplet and the three singlet states. From the plateau widths all 
characteristic energy scales can be deduced. 

\begin{figure}
\centerline{\includegraphics[width=5cm, angle=270]{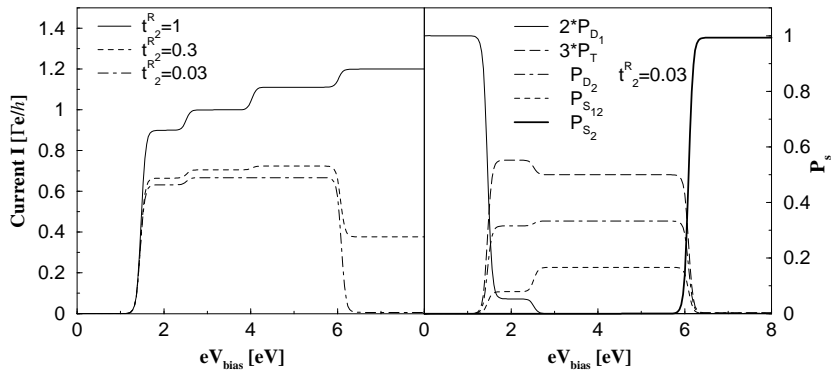}}
\caption {Left Panel: I--V--characteristics for various coupling
$t^R_2$. A pronounced NDC effect is observed for reduced $t^R_2$.
Right Panel: Occupation probability $P_s$ of the relevant molecule states
for $t^R_2=0.03$. The fat solid line indicating $P_{S_2}$ reaches 
nearly unity in the blocking regime at bias $V_{\mbox{bias}} > 6$.
We multiply the probabilities with the corresponding degeneracy,
so the sum of the $P_s$ adds up to unity.
}
\label{red2}
\end{figure}

Strong NDC behaviour is observed if one MO couples much more weakly 
to the right side than the other MO, e.g. $t^R_{2}= 0.03 ;\;
t^L_{1}=t^R_{1}=t^L_{2} =1$, see Fig.~\ref{red2}. 
For a certain bias region the current is suppressed by a factor 
$(t^R_{2})^2$. The reason for
this current decrease is the occupation of a molecule state ($S_2$ in
this case) from which the molecule can not escape anymore due to a
combination of blocking Fermi sea, Coulomb blockade and the small
coupling of an MO to the electrode.

Initially, the molecule is singly occupied in state $D_1$.  The
current starts at a bias when the first two--electron state (triplet)
becomes occupied (the ``empty'' state has higher energy for the given
parameters). The current can flow via sequential hops through MO$_1$.
The electron on MO$_2$ is essentially stuck since its tunnelling time
to the right reservoir is suppressed by a factor $(t^R_{2})^2$, and
tunnelling to the left is suppressed because of the blocking Fermi sea.

But at larger bias the electrons tunnelling onto the molecule from the
left can also form the state $S_2$, with both electrons in MO$_2$ as
depicted in Fig.~\ref{en_ivnew}.  No other electron can enter the
molecule at this bias because of the charging energy.  Since the
relaxation due to the boson coupling is very slow, the only relevant
decay of this state is via the small coupling to the right electrode.
Consequently, the molecule is stuck for a long time in state $S_2$. As
the right panel of Fig.~\ref{red2} shows the average probability
$P_{S_2}$ is nearly unity.  A relative suppression of $t^R_{2}$ by 0.3
is sufficient to achieve a pronounced NDC effect.  Increasing the
temperature will broaden the plateau steps and shift the current
maximum slightly to larger bias (not shown).  At much larger bias (not
shown), states with an additional electron become occupied, and the
current rises again.
\begin{figure}
\centerline{\includegraphics[width=5cm, angle=270]{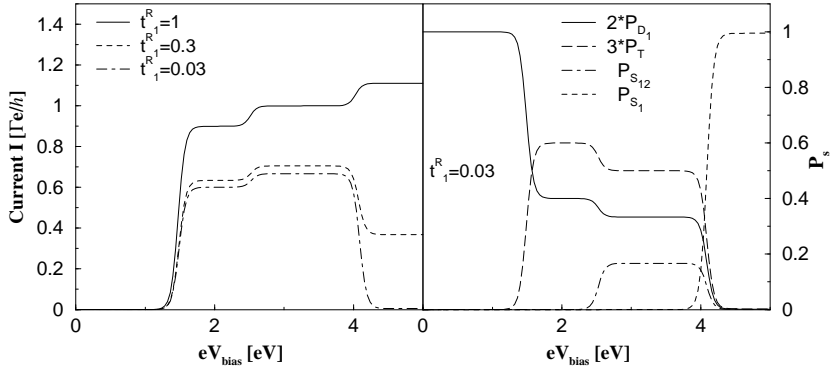}}
\caption  {Left Panel: I--V--characteristics for various coupling
$t^R_1$. Again, NDC is observed for reduced $t^R_1$, this 
time involving ${S_1}$ as the blocking state.
Right Panel: Occupation probability $P_s$ of the relevant molecule states
for $t^R_1=0.03$. The dashed line indicating $P_{S_1}$ is
nearly unity at bias $V_{\mbox{bias}} > 4$.
}
\label{red1}
\end{figure}

The left panel of Fig.~\ref{red1} shows that for the same set of
energy parameters NDC is observed also if $t^R_1$ is suppressed
instead of $t^R_2$. In this case the blocking state is $S_1$ as
indicated by the occupation probabilities in the right panel of
Fig.~\ref{red1}.  

\begin{figure}
\centerline{\includegraphics[width=5cm, angle=270]{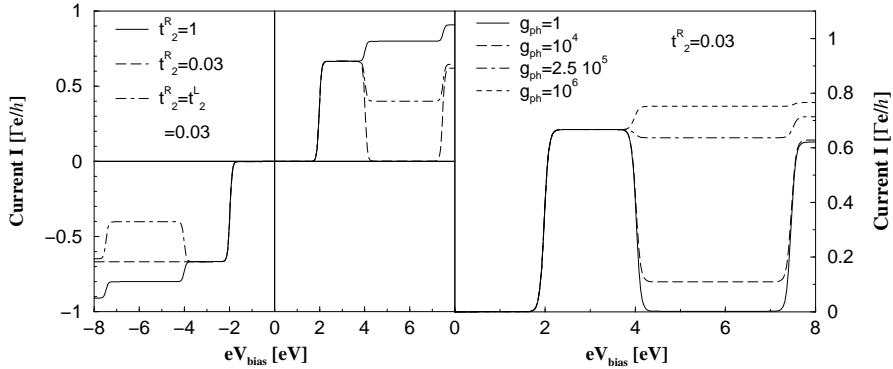}}
\caption {I--V--characteristics of the initially neutral molecule.
$\epsilon_1=-0.5$, $\epsilon_2=0.5$, $U=1.5$, $\Delta_{ex}=0.5$ and
$E_C=1.5$.  
Left Panel: NDC is observed for $t^R_2=0.03$ involving $D_2$ as 
the blocking state. If both $t^R_2$ and $t^L_2$  are suppressed
the size of the NDC is limited (see text). Right Panel: Relaxation
by photon emission $g_{ph}$ destroys NDC, but only if the coupling $g_{ph}$
is increased by several orders of magnitude over the dipole approximation.
}
\label{emp_phot}
\end{figure}

Case (2): NDC is also observed if we start from an initially
uncharged molecule, see the left panel of Fig.~\ref{emp_phot}. 
Note that NDC is observed for $t^R_2=0.03$ for positive
bias only, whereas for negative bias there is a simple step (dashed
curve).  This is because if the bias is negative MO$_2$ and
consequently $D_2$ will not become occupied at all, so the current
through $D_1$ continues to flow. We also want to point out that if
both the left {\it and} right coupling of an MO is reduced, NDC is
observed for both signs of bias, however, it is limited to a Peak to
Valley Ratio (PVR) of 2\cite{weis}, whereas the PVR for the one sided
suppression is limited by 1/$(t^R_2)^2 \gg 1$.

In the right panel of Fig.~\ref{emp_phot} we show the influence of
internal molecule relaxation by increasing the boson coupling
$g_{ph}$ for case (2) (case (1) shows similar behaviour). 
An increase in the bosonic relaxation rate by {\it six orders
 of magnitude} over the one obtained in dipole approximation is
necessary to completely eliminate NDC behaviour. It is debatable
whether coupling to vibrations of the molecule can provide such a
rate.  However, even in a situation where the coupling $g_{ph}$ is
nominally large there can be selection rules that prevent decay of
certain states. An example would be the inhibition of (direct)
transitions between states of different total spin, i.e
singlet--triplet transitions. Then, NDC will take place if the
triplet states ($T$) are lower in energy than the singlets ($S_1,
S_{12}, S_2$) {\it and} $S_1$ is the singlet of lowest energy,
different from the situation shown in Fig.~\ref{en_ivnew}.  The
blocking state is $S_1$, the singlet of lowest energy. Because of the
energy balance ``decay'' (emission of a boson) of this state could
only involve the triplets. But this is ``forbidden'' by the different
total spin of the triplets.  Therefore, as long as there is no
absorption of bosons to speak of, this scenario of NDC is stable even
in the presence of a boson coupling. The rate of absorption processes
will strongly depend on the number of bosons present. The number of
bosons increases with temperature, therefore the NDC effect will in
general decrease as the temperature is increased due to fact that the
molecule can escape the blocking state by absorption of bosons.

The proposed scenarios for NDC are  possible  explanations for a recent
experiment \cite{chen-etal} showing a PVR
of more than 1000 at $T \sim 100 K$. But in that experiment one is probing 
transport through a whole array of molecules and the nature of one 
contact side is unknown. The temperature dependence is strong 
($PVR \sim 1.5$ at room temperature) and shows a peak shift to smaller bias
with increasing temperature. 
Whereas the decrease of the PVR could be understood by the increased
absorption of bosons (see discussion above) 
the shift is without the reach of the considered model.

\begin{figure}
\centerline{\includegraphics[width=11cm]{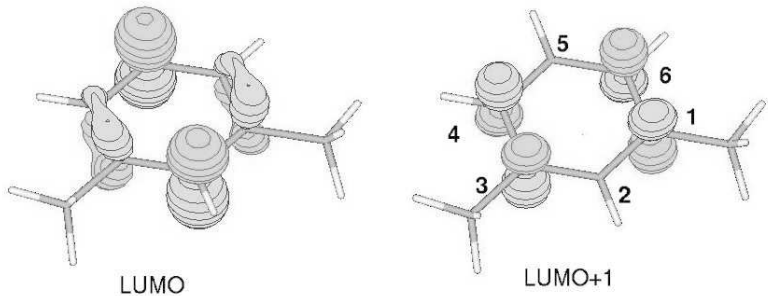}}
\caption {The LUMO and LUMO+1  for a double methyl substituted
benzene. By (anti)-symmetry, one of the MOs will have no coupling at the 2
position. 
}
\label{lumos}
\end{figure}
{\it Physical Realization.} The energetic arrangement of two closely
spaced molecular orbitals that couple very differently to the left and
right electrodes can be realized in principle with very simple
aromatic molecules that are suitably substituted to break their full
symmetry. As an example, Fig.~\ref{lumos} shows the LUMO and LUMO+1 for
a 1,3-dimethyl benzene (meta-xylene) that are closely spaced
energetically and that couple very differently on various possible
unsubstituted contact sites.
There will be MOs which are antisymmetric with respect to the 
2,5-axis mirror symmetry with vanishing wave function amplitude at the
2 and the 5 position. In contrast, the symmetric MOs will
in general have non-vanishing wave function at these positions.
If one couples the molecule at the 2 and 6 positions to electrodes, 
the LUMO couples to both electrodes
whereas the LUMO+1 would have no coupling to the electrode 'connected'
to the 2 position\cite{note_coupling}. Thus, the situation of a 
strongly MO and electrode dependent coupling seems to be generic
for small aromatic molecules with ligand groups.\\

{\it Conclusions.} We have developed a model of non--linear charge
transport through a metal--molecule--metal nanojunction. We have shown that
the interplay of charging effects and the spatially non--trivial
electronic structure of the molecule can lead to current peaks and
strong negative differential conductance. NDC will be observed no
matter whether the tunnel coupling of the lower or the higher MO is
suppressed.  For a coupling to photons in dipole approximation the
relaxation rate induced by the photons is several orders of magnitude
too small in comparison to typical tunnelling rates to have an effect.
We believe that the model is sufficiently generic to be realized in
certain classes of aromatic molecules with tunnel contacts to
electrodes.

\indent
{\it Acknowledgments.} 
The authors gratefully acknowledge 
discussions with R. Ahlrichs, D. Beckmann, T. Koch, M. Mayor, J. Reichert,
G. Sch\"on, H. Weber and F. Weigend, the financial support from
the Deutsche Forschungsgemeinschaft (DFG, WE 1863/8-1) and the 
von Neumann Center for Scientific Computing.


\begin{thebibliography}{99}
\bibitem{excitation} 
{Excited states can also be created from the 'ground state' by this coupling.
Intense irradiation of the molecule will destroy  NDC in all scenarios.}
\bibitem{note_coupling} 
{More realistically, we can say that the coupling of the antisymmetric 
MO will be much suppressed as compared to the symmetric MO. That is
all that is necessary for the effect we present.}
\end{thebibliography}
\end{document}